# Title: The Role of Driving Energy and Delocalised States for Charge Separation in Organic Semiconductors

**Authors:** Artem A. Bakulin[1]†, Akshay Rao[1], Vlad G. Pavelyev[2], Paul H.M. van Loosdrecht[2], Maxim S. Pshenichnikov[2], Dorota Niedzialek[3], Jérôme Cornil[3], David Beljonne[3], Richard H. Friend[1]*

**Affiliations:**

[1]Cavendish laboratory, University of Cambridge, JJ Thomson Avenue, Cambridge CB30HE, UK

[2]Zernike Institute for Advanced Materials, University of Groningen, Nijenborgh 4, 9747 AG Groningen, The Netherlands

[3]Laboratory for Chemistry of Novel Materials, University of Mons, Place du Parc 20, B-7000 Mons, Belgium

*Correspondence to: rhf10@cam.ac.uk

†Current address: AMOLF, Science Park 104, 1098 XG Amsterdam, The Netherlands

**Abstract:** The electron-hole pair created via photon absorption in organic photoconversion systems must overcome the Coulomb attraction to achieve long-range charge separation. We show that this process is facilitated through the formation of excited, delocalized band states. In our experiments on organic photovoltaic cells, these states were accessed for a short time (<1ps) via IR optical excitation of electron-hole pairs bound at the heterojunction. Atomistic modelling showed that the IR photons promote bound charge pairs to delocalised band states, similar to those formed just after singlet exciton dissociation, which indicates that such states act as the gateway for charge separation. Our results suggest that charge separation in efficient organic photoconversion systems, occurs through hot-state charge delocalisation rather than energy-gradient-driven intermolecular hopping.

**One Sentence Summary:** Organic photoconversion systems use delocalisation of charge carriers in the hot state to achieve efficient charge separation.

**Main text:** In contrast to the elegant photosynthetic apparatus evolved by nature,(*1*) organic photovoltaic (OPV) cells use a single heterojunction between two semiconductors to generate charge.(*2*) These semiconductors, referred to as the donor (D) and acceptor (A), are cast from solution or vacuum sublimed to form a thin film with nanoscale domains of relatively pure 'bulk' materials and large interfacial regions. This architecture, known as the bulk heterojunction,(*3*) allows for absorption of light in the bulk domains to form excitons followed by charge generation at the interface where offsets between the lowest unoccupied molecular orbitals (LUMOs) and highest occupied molecular orbitals (HOMO) of the D and A enable electron transfer, thereby dissociating the exciton.

The various conjugated macromolecules used in this study are shown in Fig. 1, along with their reported external quantum efficiencies (EQE), under short circuit conditions, which vary from high (>70%) to low (<10%) (*4-9*). This variation in performance cannot be explained simply by energy-level offsets, as these are sufficient in all listed systems to enable exciton dissociation, nor can they be explained by non-ideal charge transport.(*10, 11*) Instead, fundamental differences in electronic structure of the heterojunctions formed between materials effect the efficiency of photoconversion.

The current understanding of the photon-to-charge conversion in OPV devices is illustrated in Fig. 2A-C (*12*). Absorption of a photon produces an intramolecular singlet exciton at the D-A interface either directly or after exciton diffusion from the 'bulk' domains. The exciton then undergoes ultrafast quasi-adiabatic charge transfer, yielding the hole on D and electron on A. Having overcome the intramolecular Coulomb attraction (holding together the singlet exciton), via the use of offset energy levels, the charges remain electrostatically attracted across the D-A interface.(*10, 13*) Because of the interaction, the nature of produced state may differ from that of a free well-separated electron-hole pair (SC), so hereafter we refer to the species formed immediately after charge transfer as 'charge-transfer' (CT) states, without consideration of whether these states eventually dissociate into free carriers or stay bound until recombination.(*14*) When electron and hole do localise, so that they are immediately adjacent on either side on the heterojunction, Coulombic binding energies of several hundred meV are expected and observed.(*15, 16*) In photosynthesis, separation of charges is achieved with the aid of cascaded energy levels and ion screening.(*1, 17*) OPVs in contrast posses none of these elements, yet surprisingly, some OPV systems work very efficiently (figure 1). Hence, the fundamental question of how the organic heterojunction enables efficient long-range charge separation remains unanswered.

Here we address this issue by applying an electro-optical pump-push experiment, performed on working OPV diodes (figure 2D). A visible 'pump' pulse illuminates the OPV and creates a population of CT/CS states, as illustrated in figure 2A,C. The pump flux is kept low to give excitation densities ~$10^{16}$cm$^{-3}$, comparable to continuous solar excitation.(*18*) The charge carrier on the polymer causes a geometrical rearrangement of the chain, known as polaron formation. Polaron formation lowers the energy for charge storage by pulling two band states (π and π*) into the gap, creating the localised polaron illustrated in figure 2B.(*19, 20*) For the hole polaron, the lower gap state is singly occupied, and the optical transition (P1) between it and the π valence band is strongly allowed, forming an absorption band typically between 0.1 and 0.5eV,(*20, 21*) with the energy being determined by the degree of polaron delocalisation.(*22*) As we show below, polaron formation plays a key role in limiting long-range charge separation. In our experiments, we address the dynamics of electron-hole pairs still localised at the interface. After charge transfer, the $CT_n$ states formed are 'hot' states with energy 0.3-1 eV in excess of the lowest lying $CT_0$ states. The charge pair then relaxes through the manifold to the lowest lying $CT_0$ states giving rise to a red-shifted CT emission. (*23*) Here we selectively excited the P1 ($CT_0$-$CT_n$) transition with an IR 'push' pulse, re-populating 'hot' states that the system samples immediately after the charge transfer. The perturbation of the charge dynamics on the molecular level affected the OPV performance and the corresponding change of the photocurrent (δPC) induced by the IR-'push' pulse is recorded as a function of the delay between pump and push.(*24, 25*) In a complementary all-optical experiment (figure 2E), the effect of the push pulse on charge dynamics is observed using an additional IR 'probe' pulse.(*26*) To boost the signal and ensure that a large number of charges were excited by the push pulse, a higher excitation density ($\sim 10^{19}$cm$^{-3}$) was used in the all-optical experiment.

The push-induced states were similar to those populated directly after exciton-dissociation. The geometry of the polymer was different in the case of the 'push' experiment because the chain had relaxed to form a polaronic state. However, as established earlier (*19*), the presence of the polaron distortion leaves all of the other π band states unchanged, as is also confirmed by the calculation presented in figure S3 (*27*). This allowed us to generalise the results of our experiments to photoconverstion under solar-illumination conditions.

Figure 3A shows pump-push photocurrent data for an MDMO-PPV/PC$_{70}$BM solar cell. When the push pulse arrives after the pump, the photocurrent manifestly increased. The rise time of this increase was instrument-limited (200fs), indicating that the push interacted with species created directly after excitation. The magnitude of the push-induced photocurrent decayed slowly (~600 ps) as the delay time increased. We attribute this decay to the CT state lifetime, in agreement with previous time-resolved photoluminescence measurements.(*28, 29*) From measurements of the pump-induced absorption at the push wavelength and the flux of the push pulse, we calculated the increase of the photocurrent per IR photon absorbed per charge pair. For an MDMO-PPV/PC$_{70}$BM cell, we observed a ~5% change in relative photocurrent i.e. every IR photon absorbed by an electron-hole pair had a 5% chance of helping to dissociate it. The fast rise-time (200fs), ruled out charge trapping after diffusion to possible defect sites. We observed a longer rise of the signal for blends of P3HT and PCPDTBT, which was probably associated with delayed formation of bound states preceded by the exciton diffusion.

An important question with regard to charge separation is the role of 'excess energy' that the hot CT$_n$ state inherited from the exciton. It has been argued that this energy, which is lost during relaxation to the 'cold' CT$_0$ state, facilitates long-range charge separation.(*11, 30*) However, this interpretation has been questioned(*31*) and the role of excess energy remains ambiguous.(*14, 32-35*) We elucidated the effect of excess energy on charge separation by using below-gap excitation (figure 3A, open circles).(*31, 36*) Through the below-gap transition, CT states were populated directly at the D-A interface, rather than through singlet exciton dissociation. Although the cross-section of the S$_0$→CT transition was low, it allowed for charge generation while bypassing the 'hot' CT state. We note that this excitation was not necessarily to the 'cold' CT state, but occurs with ground state D and A geometries. Subsequent geometrical relaxation formed polaron states, giving Stokes' shifted (~0.6eV) CT luminance.(*37*) Although the CT state excited had ~0.5eV less 'excess energy' than the state created after exciton dissociation, δPC/PC kinetics (figure 3A) were similar for sub- and above-gap excitations, demonstrating that charge separation effectively occurred without the need for the large excess energy associated with the exciton.(*31*)

Figure 3B shows pump-push transients corresponding to above-gap excitation for a range of materials systems. The relative amount of additional photocurrent stimulated by re-excitation was very dependent on the particular material. Polymer-polymer blends, PFB/F8BT and P3HT/F8TBT, displayed a large push-induced increase of photocurrent (10-50%). Consistent with previous reports, the CT state lifetime in these materials was up to a few 10s of ns.(*38*) Polymer-fullerene cells demonstrated a modest photocurrent increase (<6%) and sub-ns CT-state lifetimes. Finally, the highly efficient PCDTBT/PC$_{70}$BM system showed a minor decrease of photocurrent upon IR irradiation, possibly the result of bimolecular effects.

We note that both free hole polarons (SC) and those bound within CT states may absorb the IR 'push' pulse. Although recent studies have indicated that the photoinduced absorption at 0.5 eV is related to CT states, (*39, 40*) IR-push photons can be also absorbed by weakly bound charges which would contribute to the PC even without the application of the push pulse. Thus, systems with lower EQE, where the amount of bound charges is high, demonstrate greater δPC/PC response. In contrast, in the efficient photoconversion systems, where the number of CT states is reduced, the push-induced photocurrent is low. After normalisation on the estimated amount of bound charges, we found that for all materials, the efficiency of charge separation from CT state after the 'push' was lower (roughly 50%) but still comparable to the efficiency of charge separation after the original pump pulse. This result indicates that the 'push' pulse took those excitations that had relaxed to bound CT states back to states similar to the early-time hot states formed after the singlet exciton dissociation, giving them a second chance to dissociate.

To monitor the effect of the 'push' pulse on the charges directly, we used an all-optical pump-push-probe technique (figure 2E). Figure 4a shows isotropic pump-probe transients in a MDMO-PPV/$PC_{70}BM$ film with and without the push pulse. Pump pulses created CT/SC states in the OPV material and their evolution was monitored with the IR 'probe' pulse. When the probe arrived after the pump, it was partly absorbed by the pump-generated carriers. The red curve in figure 3C reflects the population of all charged ($CT_0$/SC) states as a function of time. The green circles present analogous dynamics when a push pulse, ~2ps after the pump, brought a substantial fraction (~10%) of the charges back to the 'hot' state $CT_n$ (figure 2B,C). This re-excitation was observed as a sudden drop of signal amplitude because fewer charges were available to absorb the probe beam. The signal then re-joined the red curve as the $CT_n$ states relax to $CT_0$. Figure 3E presents the difference in response with and without the push pulse, for MDMO-PPV and PCPDTBT mixed with $PC_{70}BM$. A fast sub-100 fs component accounted for ~80% of the signal decay, with the residual relaxation of ~20% of states occurring on a longer but still relatively fast 1-20ps time scale.

Figure 3D presents the anisotropic component of the same pump-probe transient in an MDMO-PPV/$PC_{70}BM$. The initial anisotropy was lower than the 0.4 limit for random distribution of dipoles, indicating that there may be (excitonic) dynamics occurring before charge generation and/or below our time ~70fs resolution.(*34*) However, the non-zero anisotropy observed indicates that the P1 transition dipole of charges partly inherited polarisation memory from the singlet excitons. As charges moved from the chain on which they are created to another chain, the transition dipole moment associated with the charge acquired a different orientation, which was observed as depolarisation.(*34*)

Figure 3F,G shows the relative difference between 'with-push' and 'no-push' anisotropic transients. As a general trend, we observed that anisotropy was reduced by the 'push'. The anisotropy within the pulse-overlap region dropped, probably because of the different orientation of the P1 transition dipoles in the excited state. Anisotropy dynamics at longer (>0.2ps) delays displayed bi-phasic behaviour. Anisotropy was lost after the push and decreases further in following 10-50ps. The loss in anisotropy after the push corresponded to the re-excited 'hot' CT state re-localising to the neighbouring CT/SC state with a different orientation of charge-associated dipole.(*13*) The continued loss of anisotropy provides further evidences that charges became more mobile after the push pulse.

The fast charge relaxation and re-localisation observed in 3-pulse experiments are not consistent with the currently dominant models of charge separation through relaxation-assisted intermolecular hopping. However, the observed results can be explained by hole delocalisation achieved via band states. These states are formed after singlet exciton dissociation, before polaron formation occurs. Here we re-populate these states through the optically-allowed IR transition. The absorption of the IR photon promotes an electron from the delocalised valence band to the localised HOMO, instantaneously delocalising the hole, as illustrated schematically in figure 2B. Importantly, such delocalisation would then play a critical role in overriding the Coulomb binding energy.(*41*)

To confirm this hypothesis we resorted to atomistic 'many-body' modelling. We first demonstrated that optical excitation from the singly (positively) charged ground state to the lowest dipole allowed electronic state, namely the $CT_0 \rightarrow CT_n$ (P1) electronic transition in the simplified one-electron picture above, resulted in an increased intrachain hole delocalization in isolated chains of archetypical conjugated polymers (*27*). By combining force-field geometry optimization with quantum-chemical excited-state calculations of D-A pairs, we then inferred the nature of the electronic states at representative polymer-PCBM and polymer-polymer heterojunctions. Figure 4A features the charge density distribution as calculated in the lowest charge-transfer electronic excited state ($CT_0$) of a P3HT/PCBM heterojunction, while figure 4B shows the analogous situation for a higher-lying, strongly dipole coupled to $CT_0$, excited state, $CT_n$, prepared by absorption of an IR photon from $CT_0$. The P3HT-hole wavefunction, which was confined close to the PCBM electron in the $CT_0$ state, was more delocalised along the polymer backbone in the $CT_n$ state, resulting in an increased intermolecular average electron-hole separation. The electron density on the fullerene was also changed upon the excitation to the $CT_n$ state, however, as the electron is already delocalised on the fullerene molecule, this effect is not very pronounced. Nonetheless, current calculations cannot exclude further delocalization over more than one fullerene molecule when such molecules are available. Similar results were obtained for the P3HT/F8TBT interface (figures 4C,D). Thus, at both heterojunctions, the strong IR transition from the $CT_0$ to the $CT_n$ excited state, which correlates with the valence→n transition in a single P3HT chain, delocalises the hole wavefunction along the donor chain and thereby promotes a larger intermolecular electron-hole separation. Although this transient delocalisation is short-lived (sub-ps), as measured by the all-optical technique in figures 4B,D, it allows for charges to decouple and move apart at longer time scales, 10-50ps.

Although the general trend of delocalisation of charge by optical excitation is observed for both modeled systems, the extent of delocalisation is material dependent. Moving from F8TBT to PCBM increases the average electron-hole separation by 50%. This, in turn, induces even larger variations in CT-state binding energy and dissociation probability. The results of these variations are clearly seen in figures 1 and 3B. It is this difference in delocalisation of charges that causes OPV systems to demonstrate dramatically different quantum efficiencies.

In conclusion, we propose that the driving energy for charge separation in organic photoconversion systems is the energy needed to reach delocalised band states, which are critical for long-range charge separation. Although the delocalised states are extremely short-lived (<1ps), they enable charges to override the otherwise dominant Coulomb interaction. In contrast, large band offsets are not crucial for free-charge formation. Our results provide a new framework to understand charge generation in organic systems and outline the basis for the design of improved OPVs. In particular, those materials that support delocalised charge wave functions and have low reorganisation energies due to structural rigidity and suppressed torsion relaxation, should be targeted for the next generation of OPVs. This approach would mitigate the problem of polaron formation and allow for efficient charge separation with minimal band offsets, greatly increasing the open-circuit voltage and efficiency of OPVs. The superior performance of fullerene-based OPVs is explained by meeting the outlined criteria, as is the performance of recently reported high-efficiency OPVs based on porphyrins.(*42, 43*)

**Acknowledgements:** We thank D.Y. Paraschuk, J.R. Durrant, J. Clark and T. Strobel for useful discussions. A.A.B. acknowledges a Rubicon Grant from the Netherlands Organization for Scientific Research (NWO), co-financed by a Marie Curie Cofund Action. A.R. thanks Corpus Christi College for a Research Fellowship. J.C. and D.B. are research fellows of the Belgian National Fund for Scientific Research (FNRS). This project was supported by the Engineering and Physical Sciences Research Council.

**Authors contributions:** A.A.B. initiated the research; A.A.B. and A.R. fabricated the samples; A.A.B., A.R., V.G.P. and M.S.P., performed the experiments; A.A.B., A.R., M.S.P., P.H.L. and R.H.F analysed the data; D.N., J.C. and D.B. performed quantum-chemical simulations; A.A.B., A.R. and R.H.F. wrote the paper; all the authors contributed to the discussion of the results and to the manuscript preparation.

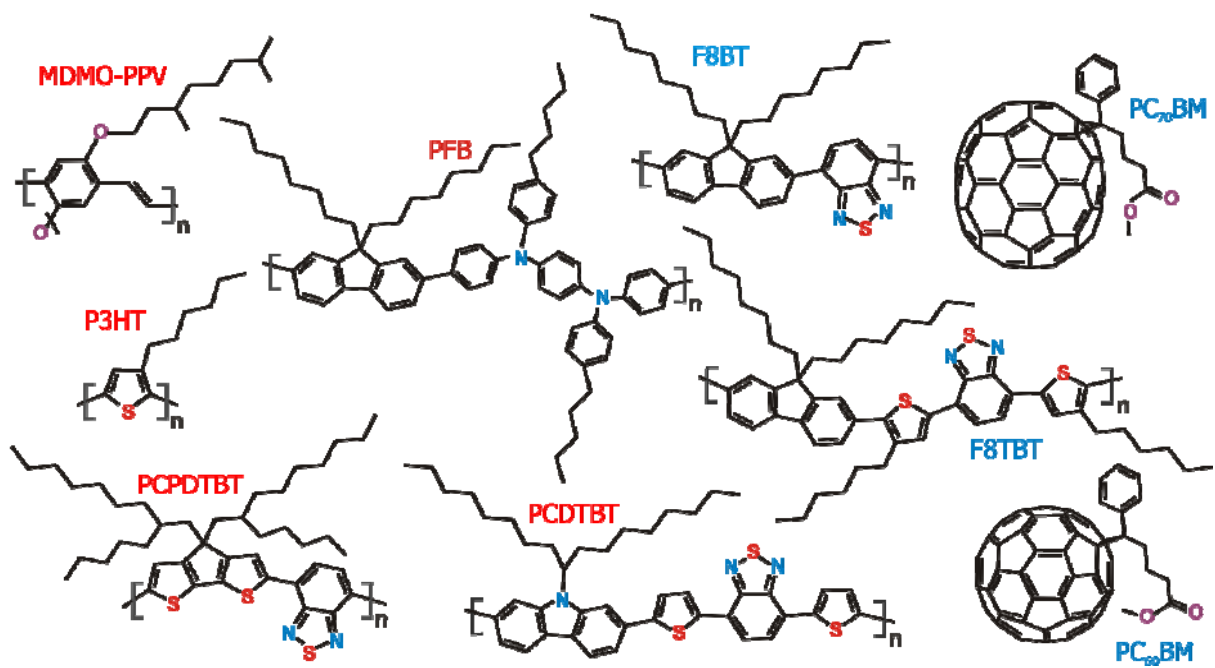

**Fig. 1.** Chemical structures of donor (left, red captions) and acceptor (right, blue captions) materials utilised in this study. Typical external quantum efficiencies of the corresponding OPVs (4-9), defined as the number of collected electrons collected per incident photon are: PFB:F8BT - 6%; P3HT:F8TBT - 25%; MDMO-PPV:$PC_{70}BM$ - 40%; PCPDTBT:$PC_{70}BM$ - 50%; P3HT: $PC_{60}BM$ - 70%; PCDTBT: $PC_{70}BM$ - 80%.

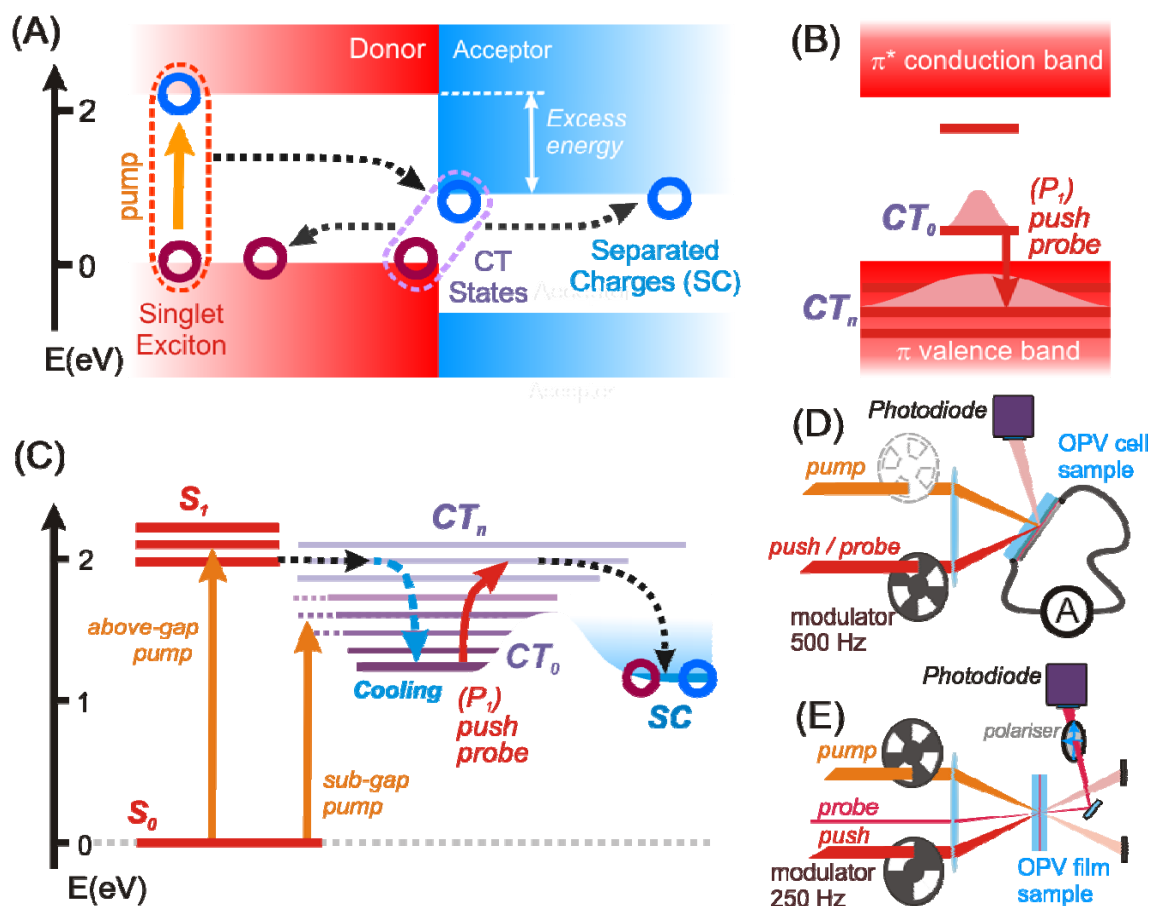

**Fig. 2. (A)** Band diagrams for a typical OPV and **(B)** for cationic state on the polymer donor. **(C)** Free-energy state diagram of the same OPV system. Singlet, Charge-Transfer (CT, lowest-lying - $CT_0$, band states - $CT_n$) and separated-charges (SC) states are shown; positive charge density distribution in **(B)** is indicated by pink contour. Solid arrows show optical transitions and dashed arrows indicate energy/charge transfer pathways involved in photoconversion. Layouts of **(D)** pump-push photocurrent and **(E)** 3-pulse transient-anisotropy experiments.

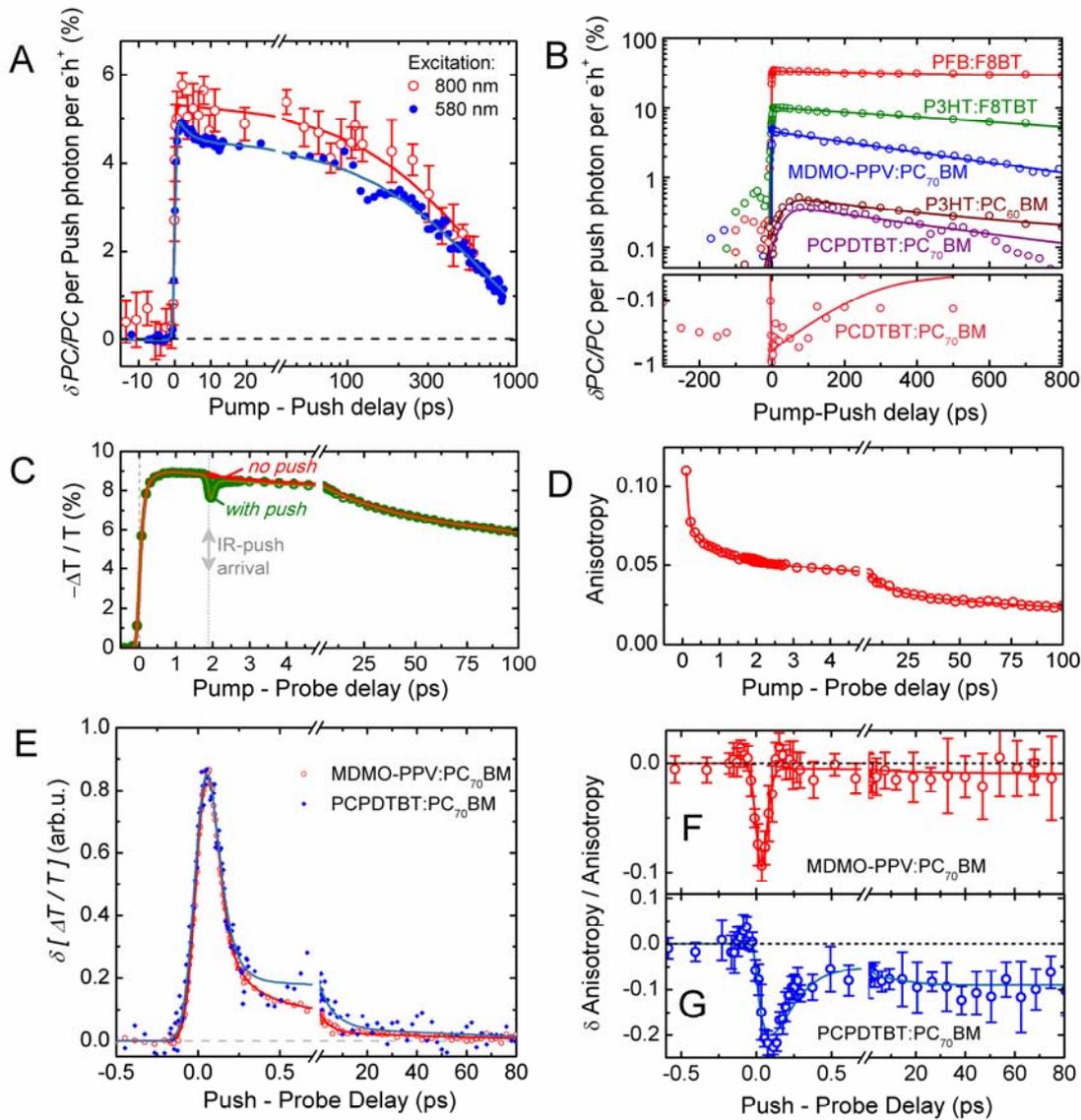

**Fig. 3.** The results of ultrafast spectroscopy experiments. **(A)** Pump-push photocurrent transients for MDMO-PPV/PC$_{70}$BM cell. Scatters show increase in relative photocurrent induced by the push pulse. Pump photons had energy above (580 nm) and below (800 nm) polymer and fullerene band gaps. **(B)** The results of pump-push photocurrent experiments on a set of OPVs under above-gap excitation. **(C)** Transient-absorption and **(D)** transient-anisotropy kinetics for MDMO-PPV:PC$_{70}$BM film excited at 580 nm and probed 3 μm. Green line in figure **(C)** shows dynamics with push pulse arriving at ~2 ps delay. The effect of push pulse on the isotropic **(E)** and anisotropic **(F,G)** transients, calculated from the difference between 'with-push' and 'no-push' measurements, for MDMO-PPV/PC$_{70}$BM, PCPDTBT/PC$_{70}$BM films. In all figures, solid lines are (multi)exponential fits convoluted with the Gaussian function at zero delay.

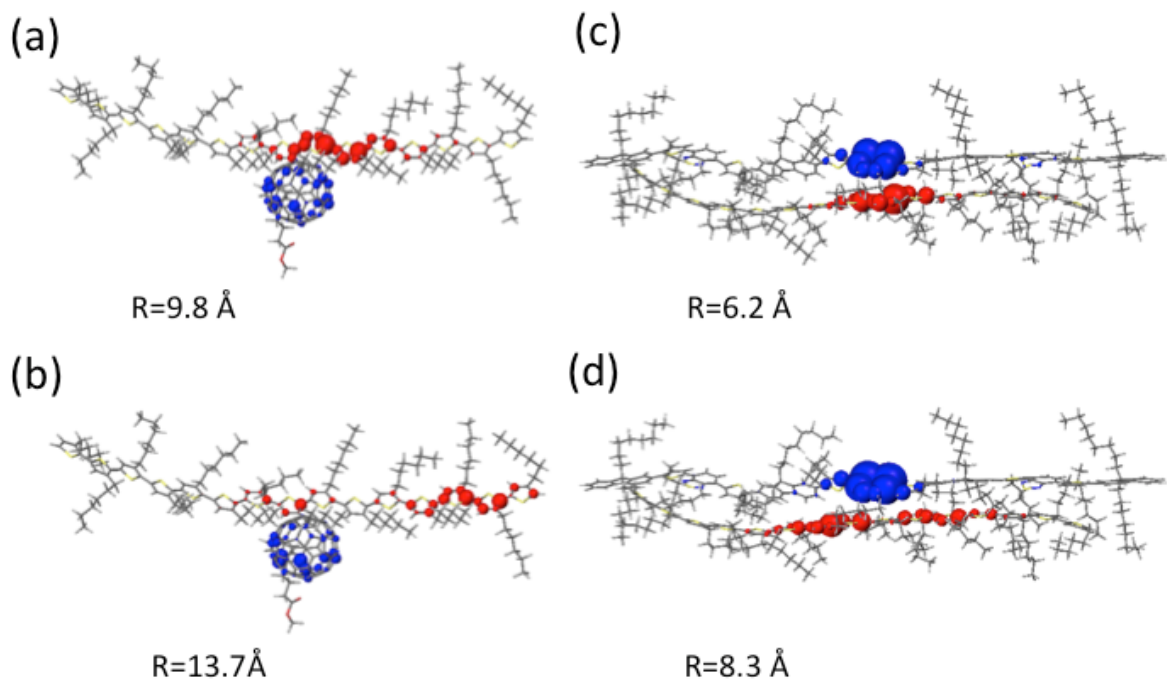

**Fig. 4.** Microelectrostatic simulations of the charge distribution at the P3HT/PCBM **(A,B)** and P3HT/F8TBT **(C,D)** heterojunction, with electron and hole densities shown in blue and red respectively. **(A,C)** show the charge distribution in the lowest charge-transfer state configuration, and **(B,D)** the excited charge-transfer state configuration created upon absorption of an IR-'push' photon. R stands for the average electron-hole separation.